\documentclass[prl,twocolumn,superscriptaddress,showpacs,nofootinbib]{revtex4}

\usepackage{graphicx}

\usepackage{bm}
\usepackage{amssymb}
\usepackage{epstopdf}
\usepackage{amsmath}
\usepackage{amstext}
\usepackage{latexsym}
\usepackage{hyperref}
\usepackage{pstricks}
\usepackage{ulem}

\newcommand{\ket}[1]{\left\vert#1\right\rangle}

\newcommand{\beq}{\begin{equation}}
\newcommand{\eeq}{\end{equation}}

\begin{document}

\title{Entanglement Spectrum, Critical Exponents and Order Parameters in Quantum Spin Chains}
\author{G. De Chiara}
\affiliation{Departament de F\'isica. Universitat Aut\`{o}noma de Barcelona, E-08193 Bellaterra, Spain}
\affiliation{Centre for Theoretical Atomic, Molecular and Optical Physics, School of Mathematics and Physics, QueenÕs University Belfast, Belfast BT7 1NN, United Kingdom}
\author{L. Lepori}
\affiliation{Departament de F\'isica. Universitat Aut\`{o}noma de Barcelona, E-08193 Bellaterra, Spain}
\author{M. Lewenstein}
\affiliation{ICREA, Instituci\`o Catalana de Recerca i Estudis Avan\c{c}ats, E08011 Barcelona}
\affiliation{ICFO--Institut de Ci\`encies Fot\`oniques,
Parc Mediterrani de la Tecnologia, 08860 Castelldefels, Spain}
\author{A. Sanpera}
\affiliation{ICREA, Instituci\`o Catalana de Recerca i Estudis Avan\c{c}ats, E08011 Barcelona}
\affiliation{Departament de F\'isica. Universitat Aut\`{o}noma de Barcelona, E-08193 Bellaterra, Spain}

\begin{abstract}
We investigate the entanglement spectrum near criticality in finite quantum spin chains. 
Using finite size scaling we show that when approaching a quantum phase transition, the Schmidt gap, i.e., 
the difference between the two largest eigenvalues of the reduced density matrix $\lambda_{1},\lambda_{{2}}$,  
signals the critical point and scales with universal critical exponents related to the relevant operators of the 
corresponding {\it perturbed} conformal field theory describing the critical point. Such scaling behavior allows us to identify explicitly 
the Schmidt gap as a local order parameter.
 
\end{abstract}

\maketitle

The characterization of phase transitions in strongly correlated systems has been
traditionally based on the behavior of expectation values of local operators and two point correlators.
This is the spirit of the ``standard" Ginzburg--Landau scenario of second order phase transitions where many-body systems order (locally) for
given values of a ``control" parameter $g$ of the Hamiltonian. The order, which is associated to the breaking of some symmetry,  
is manifested by an order parameter, $Q$, whose expectation value in the ordered phase is different from zero while it becomes exactly
zero at criticality, i.e., for $g=g_c$. Near criticality, order parameters and other physical quantities describing the many-body system exhibit scaling behavior, i.e., a power dependence with $|g-g_c|$. The exponents of these power laws are called critical exponents. 
The scaling exponents reflect the universality class of the theory, which is independent of the microscopic details of the model but depends only on global properties such as the symmetries and dimensionality of the Hamiltonian.

Recently, an alternative approach to understanding quantum many-body
systems and their simulatability \cite{Schuch07}, exploits the
entanglement content of the ground states of
such systems. First attempts within this approach focused on the study of quantum phase transitions (QPT) in spin chains \cite{Osborne02,Osterloh02}. 
A good figure of merit to measure the bipartite correlations embedded in the ground state of a spin chain of length $L$,
is the entanglement entropy (EE). It refers to the von Neumann entropy of a block of $\ell<L$ spins,  $S=-\sum_i \lambda_{i}\log \lambda_{i}$, where the $\lambda_{i}$, are the eigenvalues of the reduced density matrices obtained after the bipartite splitting of the ground state.
Its finite size scaling, i.e., the dependence of the EE with the size of the block $\ell$, shows remarkable properties \cite{Vidal03}. At criticality, the EE diverges logarithmically
as $S\sim c\log \ell$ \cite{Holzhey94,Vidal03,Korepin04,Calabrese04,Hastings07} where $c$ is the so-called central charge of the corresponding QPT as provided by conformal field theory (CFT).
This universal logarithmic behaviour of the entanglement entropy at criticality underpins the conformal invariance of QPT in 1D, and leads also to a universal --depending asymptotically only on c--  distribution of the eigenvalues of the reduced density matrix \cite{Calabrese08}.  
On the other hand, outside but close to criticality, when the system is gapped, 
the EE scales as $S\sim c\log(\xi)$ \cite{Calabrese04} being $\xi$ the correlation length that sets the relevant scale for long-distances physics. It is in this regime when the system is not any longer entirely constricted by the central charge, and its description
depends also on the full set of Hamiltonian parameters tuned to move away from the critical point.


Thus, it is to be expected that away from criticality, further information about the many-body system that is not included in EE can be obtained from the set of parameters $\{\lambda_{i}\}$~\cite{Li08,note}. We shall denote this set as the entanglement spectrum.  
Indeed, it is known that the (topological) Haldane phases appearing 
for integer spin chains are characterized by a double (or higher) degeneracy of the entire entanglement spectrum, resulting from applying the symmetries of the Hamiltonian 
to the eigenstates of the reduced density matrix~\cite{Pollmann10}.

Here, we examine the behavior of the entanglement spectrum for finite spin chains,  of length $L$,  in the vicinity of a QPT.  For simplicity, we assume a symmetric bipartite splitting and we show that when approaching a QPT,  the difference between the two largest (nontrivially degenerate) eigenvalues of the reduced density matrix $\lambda_{1},\lambda_{{2}}$
signals correctly the critical point and scales with the universal critical exponents. We relate
this result to the relevant operators of the corresponding CFT valid to describe
perturbations from the critical QPT point.

Before proceeding further towards our main result, let us just clarify the notation used throughout. For any bipartite splitting, the ground state of the spin chain can be expressed 
by its Schmidt decomposition
$
\ket{\psi_{GS}}=\sum_i\sqrt{\lambda_i} \ket{\phi_i^{L}}\otimes
\ket{\phi_i^R}, \label{schmidt}
$ where $\lambda_i\geq 0$ are the Schmidt eigenvalues with respect
to the partition (left or right) sorted in decreasing order and $\ket{\phi_i^L}$ and $\ket{\phi_i^R}$
are the Schmidt eigenvectors \cite{note2}.
Our study refers to partitions of the spin
chain in real space and it differs from previous studies in the quantum Hall regime and some spin chains \cite{Li08,Thomale},
in which the entanglement spectrum is analyzed after a cut in momentum space is performed.

We first review the finite size effects exhibited by the entanglement spectrum at criticality. In the framework of CFT (see \cite{difrancesco,Calabrese04} and references therein) for a chain of size $L$  with periodic boundary conditions, 
the reduced density matrix of a block of size $\ell$ can be expressed as \cite{Holzhey94,Orus05}
\beq 
\rho_{\ell}=\frac{1}{Z_{\ell}(q)}q^{-c/24} q^{L_{0}} 
\eeq
where $c$ is the central charge, $L_0$ is the zero-generator of the chiral Virasoro algebra, $Z_\ell(q)=\textrm{Tr}[q^{L_{0}}] =1+n_{1}q^{\alpha_{1}}+n_{2}q^{\alpha_{2}}+... $ 
is the partition function of the subsystem 
where 
$q=\exp(i 2\pi \tau)$ with $\tau=(i\kappa)/[\log(\ell/\eta)]$, $\eta$ is a regularization cutoff and $\kappa$ a positive constant. The Schmidt gap, defined as 
$\Delta\lambda=\lambda_{1}-\lambda_{2}$, can be expressed as
\beq
\Delta\lambda(\ell,g_{c})=\frac{1-q^{\alpha_1}}{Z_{\ell}(q)}\sim \frac{1-q^{\alpha_{1}}}{\ell^{c/12}}
\label{schmidtgapCFT}
\eeq
where, in the last term, we made use of the fact that, at criticality, the largest eigenvalue of the reduced density matrix, the so-called single copy
entanglement~\cite{singlecopy}, corresponds to half of the von Neumann entropy $\lambda_1\sim \ell^{-c/12}$.
The coefficients $\alpha_{i} > 0$, with degeneray $n_{i}$, related to the scaling dimension of the
relevant operators of the theory, correspond to the eigenvalues of the operator $L_{0}$. Therefore, from CFT arguments, for finite systems at criticality the Schmidt gap is solely determined by $c$ and the smallest eigenvalue $\alpha_1$.  Notice that in the thermodynamic limit 
the Schmidt gap closes. 

 
 Here we study the entanglement spectrum $\{\lambda_i\}$ in the {\it vicinity} of a QPT
by varying a single parameter of the Hamiltonian describing the spin chain. In many known cases, this ensures an unambiguous identification of this operator with the operator content of the CFT.
In these cases, the mass gap and corresponding order parameter scale as a power-law in $|g-g_{c}|\rightarrow 0$, with certain critical exponents. The latter ones depend on the central charge and on the primary operator perturbing the CFT.


Our main result shows that the Schmidt gap develops the same scaling behavior as the one found for the mass gap and the magnetization, when approaching a QPT.  That is, when properly scaled, the quantities $\Delta\lambda(L,g)$ obtained for different chain lengths $L$ can be made to cross in a single point ---which signals $g_{c}$--- and to collapse into a single curve, with the same universal scaling exponents of the QPT. It is indeed remarkable that the information encoded in just
these two eigenvalues reproduces global properties of the ground state out of criticality, where equation (2) is not  valid anymore.

We illustrate our results by analyzing first the entanglement spectrum and the Schmidt gap in the integrable transverse field Ising model.
We move then to the longitudinal spin-1/2 Ising model and the nonintegrable uniaxial spin-1 chain which we solve numerically. 
Let us stress that at criticality, and for sufficiently long chains our results agree with those obtained by CFT.

\begin{figure}[t!]
\begin{center}
\includegraphics[scale=0.36,angle=270,clip=false, trim=3 35 0 0]{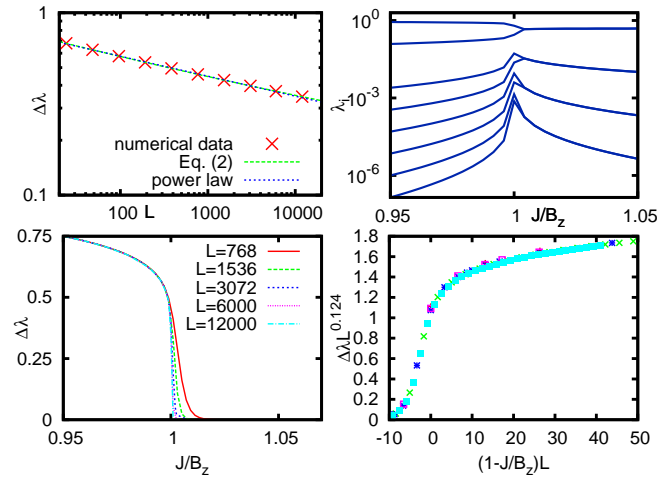}
\caption{Transverse Ising model. Top-Left: Schmidt gap at the critical point ($J=B_z$) as a function of the chain size. The numerical data (symbols) are compared to a fitting from Eq.~\eqref{schmidtgapCFT} (dashed) and to a simple power law (dotted).
Top-Right: Entanglement spectrum: the first 8 eigenvalues $\lambda_i$ are shown as a function of $J/B_z$ for $L=12000$. Bottom-Left: Schmidt gap as a function of $J/B_z$ for different lengths. Bottom-Right: FSS analysis of $\Delta\lambda$ for $L=768,1536,3072,6000,12000$. }
\label{fig:1}
\end{center}
\end{figure}

{\bf Spin1/2 Ising model-} We start with the following spin-1/2 Ising model:
\begin{equation}
\label{eq:Isingmodel}
H_{Ising}=-J\sum_i \sigma_x^i\sigma_x^{i+1} - B_z\sum_i \sigma_z^i +B_x\sum_i \sigma_{x}^{i}.
\end{equation}
where $\sigma_{k=x,y,z}$ denotes the Pauli matrices. We assume open boundary conditions.
For $B_{x}=0$, Eq.~\eqref{eq:Isingmodel} reduces to the paradigmatic transverse-field Ising model with a critical point at $J=B_z$ 
and central charge $c=1/2$.
To compute the entanglement spectrum we employ the Jordan-Wigner and Bogoliubov transformations to map the transverse Ising model into  a system of noninteracting fermions. We then proceed as in \cite{Peschel2009} to obtain the Schmidt eigenvalues $\lambda_i$ of the reduced density matrix after
cutting a chain of size $L$ (from 48 to 12000) in two symmetric halves ($\ell=L/2$). Our results at criticality, $J=B_{z}$, are summarized in the top-left panel of Fig.~\ref{fig:1}, where $\Delta\lambda$, shown as a function of $L$, is fitted by a simple power law function $L^{-\gamma}$ and by the expression given by Eq.~\eqref{schmidtgapCFT} using $c=1/2$,  $\alpha_{1}=1/8$, and  $\eta$ and $\kappa$ as fitting parameters. Both fittings are 
very accurate but when looking at the marginals of the fitting functions, Eq.~\eqref{schmidtgapCFT} matches better the Schmidt gap values, in particular
for the largest chain size, i.e. $L=12000$.

 For $J>B_z$, the order parameter $M_x =L^{-1} \sum_i\langle \sigma_{x}^i\rangle$, scales, in the vicinity of the critical point, as $M_x\sim |J/B_z-1|^\beta$ with critical exponent  $\beta=1/8$ while the correlation length, $\xi\sim  |J/B_z-1|^{-\nu}$ scales with $\nu=1$.
We study now perturbations from the critical point by changing the parameter $J/B_z$ while keeping $B_x=0$. The results for the entanglement spectrum are plotted in Fig.~\ref{fig:1} (Top-Right panel), showing a tendency of the eigenvalues to collapse at the critical point (notice the vertical logarithmic scale). 
The doublets appearing for $J>B_z$ are due to the unbroken $\mathrm{Z}_2$ symmetry of the ground state. 
The behaviour of the Schmidt gap $\Delta\lambda$ as a function of $J/B_z$ for different lengths is shown in Fig.\ref{fig:1} (Bottom-Left panel). 
The Schmidt gap closes very rapidly when $J$ approaches $B_z$. 
For finite size systems, the asymptotic behavior corresponding to the thermodynamic limit  
can be retrieved using finite size scaling (FSS)~\cite{Fisher72}.  
Near criticality, the dependence of the order parameter $Q$ with the finite size $L$ is given by
\begin{equation}
Q(L, g) \simeq L^{-\beta_Q/\nu} f_Q\left(|g-g_c| L^{1/\nu} \right),
\end{equation}
where  $\nu$ characterizes the divergence of the correlation length
while $\beta_Q$ is the order parameter critical exponent: $Q\sim |g-g_c|^{\beta_Q}$. 
To obtain critical exponents,  we change the coefficients $\mu_{1,2}$ until we observe the collapse of  $\Delta\lambda\,L^{\mu_1}$ as a function of $|g-g_c\;|L^{\mu_2}$ for all used lengths $L$ in the simulations.  In this way we extract $\beta_Q=\mu_1/\mu_2$ and $\nu=1/\mu_2$.  The results of the FSS of the Schmidt gap are depicted in Fig.~\ref{fig:1} (Bottom-Right panel).  
From the scaling of the Schmidt gap we obtain the critical exponents $\beta_{\Delta\lambda}=0.124 \pm 0.002$ and $\nu_{\Delta\lambda}=1.00\pm 0.01$. These values coincide with the critical exponents of the  Ising universality class, suggesting that the Schmidt gap in the vicinity of the critical points scales universally with the same critical exponents.  We remark that Fig.\ref{fig:1} contains results computed using the symmetric superposition of the two degenerate ground states in the ferromagnetic phase as obtained through the Jordan-Wigner transformation. The same results for the critical exponents are obtained if  the $Z_2$ symmetry is explicitly broken and only one of the two degenerate ground states is considered.
\begin{figure}[t!]
\begin{center}
\includegraphics[scale=0.36,angle=0,clip=true, trim=3 10 11 0]{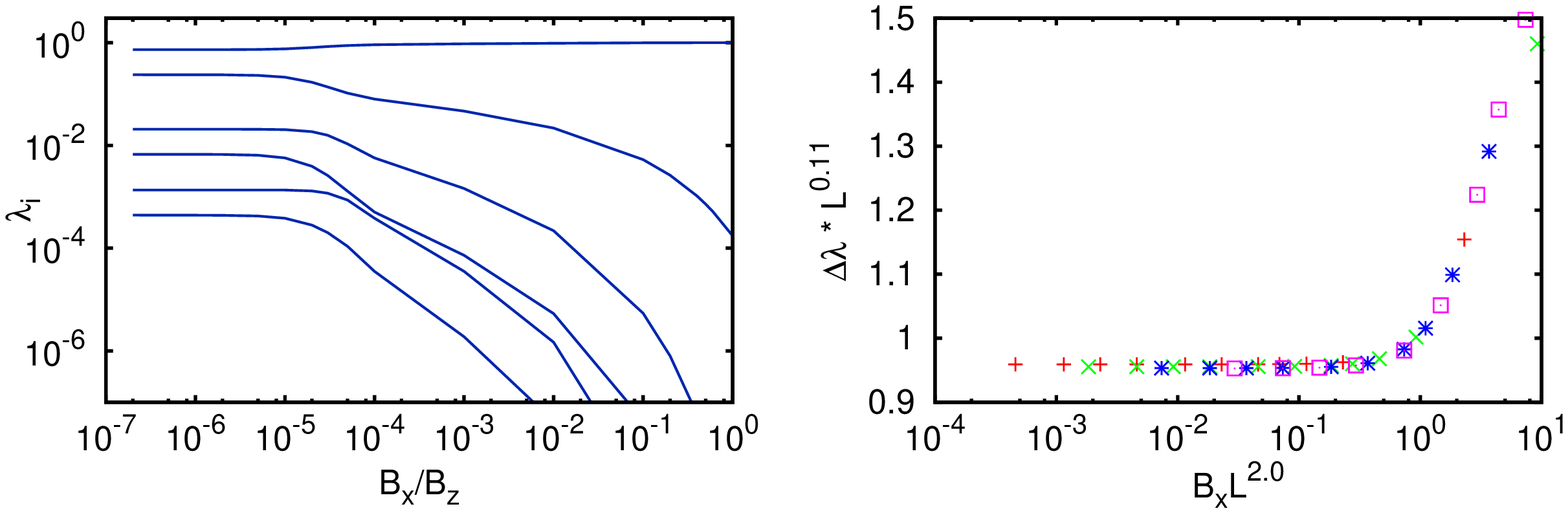}
\caption{Ising model with longitudinal perturbation, DMRG calculations. Left: The first $6$ Schmidt eigenvalues as a function of $B_x/B_z$ in logarithmic scale for $L=384$. The critical point corresponds to $B_x\to 0$. Right: FSS analysis of $\Delta\lambda$ for $L=48,96,192,384$. }
\label{fig:2}
\end{center}
\end{figure}

The surprising, yet interesting, results obtained for the transverse Ising model pushes us to look at the same critical point of the transverse Ising model 
$J=B_z$ but adding also a  very weak longitudinal field perturbation, i.e. $B_{x}\ll B_z$  in Eq. \eqref{eq:Isingmodel}.
This integrable model~\cite{Zamlodchikov89} has the same critical point as the transverse Ising model, $B_{x}=0$, and thus a central charge $c=1/2$, but belongs to a different universality class. This implies for instance that its critical exponents ($\nu=8/15$ and $\beta=1/15$) do not coincide with the exponents of the transverse Ising Hamiltonian. 
 However notice that the ratio $\beta/\nu=1/8$ is the same in both cases as predicted by scaling dimensional analysis.
We use the density matrix renormalization group (DMRG)\cite{dmrg} to find the ground state of the Hamiltonian Eq.~\eqref{eq:Isingmodel} in the vicinity of the critical point and to obtain its entanglement spectrum. The results are shown in Fig.~\ref{fig:2}. In the left panel we show the first 6 eigenvalues $\lambda_i$ which, when approaching the critical point, tend to form a band as expected from CFT \cite{Peschel87,Okunishi99}.  In the right panel we show the FSS analysis of the Schmidt gap $\Delta\lambda$ from which we infer the estimate for the critical exponents: $\nu= 0.50\pm0.05$ and $\beta_{\Delta\lambda}=0.055\pm0.005$ which are in good agreement with the exact values of the corresponding exponents, confirming that the scaling exponents of the Schmidt gap properly reflect the perturbation on the critical point.%

\begin{figure}[t]
\begin{center}
\includegraphics[scale=0.6]{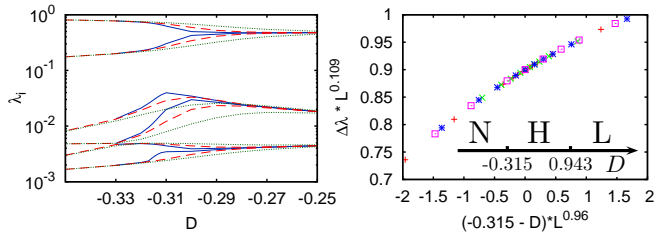}
\caption{Spin-1 model. Left: The entanglement spectrum for $\theta=0$ and as a function of $D$ nearby the N\'eel-Haldane quantum phase transition (for clarity only the first 6 eigenvalues are depicted). Dotted, dashed and continuous lines correspond to spin chains of length L=192,384 and 768, respectively. The location of the critical point obtained after FSS analysis of the Schmidt gap is $D=-0.315$.
Right: FSS analysis of the Schmidt gap together with a schematic plot of the critical points marking the transition between N\'eel (N), Haldane(H) and 
Large-D (L) phases for $\theta=0$ as a function of D.}
\label{fig:theta0}
\end{center}
\end{figure}

Finally, we consider also the nonintegrable spin-1 Heisenberg chain with uniaxial anisotropy which presents 
a very rich phase structure \cite{Rodriguez10}. The Hamiltonian describing such a system is given by
\begin{equation}
\label{eq:H} H =\sum_i H_i(\theta) +D\sum_i S_{zi}^2,
\end{equation}
where $\bm{S}_i=(S_{xi},S_{yi},S_{zi})$ are the $i$th site angular
momentum operators and $H_i(\theta)=\cos(\theta)
\bm{S}_i\cdot\bm{S}_{i+1}+ \sin(\theta)
(\bm{S}_i\cdot\bm{S}_{i+1})^2$. 
For $D=0$, the model is the well known 
bilinear-biquadratic spin-1 chain, whose phase diagram as a
function of $\theta\in[-\pi;\pi]$ is well established. 
For $-\pi/4<\theta<\pi/4$, the system is in the
aforementioned Haldane phase. A sketch of the phases surrounding the Haldane phase when the uniaxial anisotropy $D$ is present is 
depicted in the inset of Fig.~\ref{fig:theta0}.
We concentrate our analysis here in the  N\'eel-Haldane phase transition along $\theta=0$ using the uniaxial anisotropy, $D$, as the control parameter 
but the same results are obtained if the transition is crossed at a fixed value of $D$ and using $\theta$ instead as the control parameter. In the N\'eel phase, the staggered magnetization per site $M_z=L^{-1}\sum_i (-1)^i \langle S_{zi}\rangle$ is the order parameter. As before, we calculate the ground state and the Schmidt eigenvalues after a bipartite cut of the spin chain into two symmetric halves, obtain the Schmidt gap and perform FSS for different chain sizes.  In our DMRG simulations, we apply a small magnetic field to the first spin to select one of the two degenerate ground states.  This technical trick has the advantage of stabilising and speeding-up the algorithm.
  Notice  that in the thermodynamic limit only one of the two degenerate ground states is selected: our procedure allows to investigate correctly this limit.
The entanglement spectrum is shown in the left-panel of Fig.~\ref{fig:theta0} where the expected doubly degeneracy in the Haldane phase is evident.
The FSS analysis of the Schmidt gap is shown in Fig.~\ref{fig:theta0} (right-panel).
Our estimates for the location of the critical point $D_{c}$ and critical exponents $\beta,\nu$ obtained using FSS on the staggered magnetization and 
on the Schmidt gap are summarized in Table~\ref{tab:ising} and compared with those obtained using quantum Monte Carlo calculations~\cite{Albuquerque09}. The critical exponents obtained from the Schmidt gap confirm the well known fact that the N\'eel-Haldane transition belongs to the same universality class as the transverse Ising model. There is a $12\%$ discrepancy in the critical exponent $\beta$ that can be attributed to the relatively small size of the chains, as compared to the 12000 sites used in the transverse spin-1/2 Ising model. 
\begin{table}[t]
\begin{center}
\begin{tabular}{|l|l|l|l|}
\hline
Observable & $D_c$ & $\beta$ & $\nu$ \\
\hline
$M_z$ (QMC) & -0.316 & 0.147 & 1.01 \\
\hline
$M_z$ (DMRG) & -0.315 & 0.11 & 1.01 \\
\hline
$\Delta\lambda$ (DMRG) & -0.315 & 0.11 & 1.04 \\
\hline
\end{tabular}
\caption{\label{tab:ising}
N\'eel-Haldane transition in the spin-1 model. The first line contains results obtained with quantum Monte Carlo of Ref.~\cite{Albuquerque09}. The other two lines contain the FSS analysis from the order parameter (staggered magnetization) and the Schmidt gap respectively for system sizes of $L=96,192,384,768$.}
\end{center} 
\end{table}

Summarizing, we have analyzed the entanglement spectrum $\{\lambda_{i}\}$ for finite size systems near the QPT of some integrable and nonintegrable spin models. Notably, the information encoded just in the difference between its two largest eigenvalues (Schmidt gap), reproduces the scaling behavior of conventional order parameters closing at criticality and displaying universal critical exponents on its finite size scaling. 
The first result stems naturally from the fact that at criticality and in the thermodynamical limit the entanglement spectrum must become a continuous distribution of eigenvalues implying the closure of the Schmidt gap. While finite size effects generally prevent such closure, the presence of a symmetry which is preserved in the Schmidt decomposition of the ground state, like the $Z_2$ parity in the transverse Ising or in the considered transitions in the spin-1 chain \cite{note2}, forces the entanglement spectrum to organize in multiplets making finite size effects much less important. The scaling behavior of the Schmidt gap when approaching a QPT is instead less trivial and points out that different models belonging to the same universality class when perturbed with a single operator share a common Schmidt scaling behavior, while fulfilling majorization relations \cite{Orus05}. Although we have not found such scaling for other differences between Schmidt eigenvalues, we cannot rule out the finite size effects and leave this question for a forthcoming study \cite{conj}.

{\it Acknowledgements:} We thank R. Fazio, E. Berg, A. Gendiar and A. Turner
for enlightening discussions as well as the Kavli Institute of Theoretical Physics, Santa Barbara, California, USA for hospitality.  
We acknowledge financial support from: 
Spanish Goverment (JdC, FIS2008-01236;00784), Generalitat de Catalunya (SGR2009-00985;00347), European Community (ERC -QUAGATUA), EU (AQUTE, NAMEQUAM) and ERDF: European Regional Development Fund).
We used the DMRG code available at \url{http://qti.sns.it/dmrg/phome.html}.


\begin{thebibliography}{99}
\bibitem{note}
We note that sometimes in the literature the entanglement spectrum is defined by the values $-\log\lambda_i$.
\bibitem{note2}
For an infinite system with translational invariance the location of the cut does not affect the results, 
provided that the cut do not break any symmetry. 
For finite systems though, the location of the cut gives a correction to the entropy. 
When the cut breaks the symmetry, e.g. in a dimerized phase, special care has to be taken.
\end{thebibliography}
\end{document}